\documentclass[aps,prl,twocolumn,superscriptaddress,longbibliography]{revtex4-1}

\usepackage[colorlinks=true,citecolor=blue]{hyperref} 
\usepackage{graphicx}
\usepackage{bm}
\usepackage{amsmath}
\usepackage{amssymb}

\DeclareGraphicsExtensions{.png,.jpg,.eps}
\usepackage{xcolor}

\newcommand{\beq}{\begin{eqnarray}}
\newcommand{\eeq}{\end{eqnarray}}
\begin{document}

\author{Guangze Chen}
\affiliation{Department of Applied Physics, Aalto University, 02150 Espoo, Finland}

\author{J. L. Lado}
\affiliation{Department of Applied Physics, Aalto University, 02150 Espoo, Finland}

\title{Tunable moire spinons in magnetically encapsulated twisted van der Waals quantum spin-liquids}

\begin{abstract}
Quantum spin-liquid 
van der Waals magnets such as TaS$_2$,
TaSe$_2$, and RuCl$_3$
provide a natural platform to
explore new exotic phenomena associated with
spinon physics, whose properties can be controlled by exchange proximity
with ferromagnetic insulators such as 
CrBr$_3$.
Here we put forward
a twisted van der Waals heterostructure based on a
quantum spin-liquid bilayer encapsulated
between ferromagnetic insulators. 
We demonstrate the
emergence of spinon flat bands and topological spinon states in such
heterostructure, where the emergence
of a topological gap is driven by the twist. We further show
that the spinon bandstructure
can be controlled via exchange proximity effect
to the ferromagnetic leads.
We finally show how by
combining small magnetic fields with tunneling spectroscopy,
magnetically encapsulated heterostructures provide a way of characterizing
the nature of the quantum spin-liquid state.
Our results put forward twisted quantum spin-liquid bilayers as potential
platforms for exotic moire spinon phenomena, demonstrating the versatility of
magnetic van der Waals heterostructures.
\end{abstract}

\date{\today}

\maketitle

Magnetic van der Waals materials have risen as
a highly versatile family of compounds in the two-dimensional
realm\cite{Park_2016,huang2017layer,Gong2017,Gibertini2019}. These
materials attracted much research interest as their two-dimensional nature
provides a platform to electrically control magnetism\cite{Deng2018,Huang2018,Jiang2018,PhysRevLett.117.267203}, 
design magnetic tunnel junctions\cite{ghazaryan2018magnon,Klein1218,Song2018,jiang2018spin},
topological superconductivity,\cite{Kezilebieke2020,2020arXiv201109760K} and
exploit magnetism in generic
van der Waals heterostructures\cite{Zhong2017}. 
Remarkably, specific 
magnetic van der Waals materials
 such as
TaS$_2$, TaSe$_2$ and RuCl$_3$
provide a realization of an exotic phase of matter, quantum spin-liquid (QSL) states\cite{Law2017,Klanjek2017,Chen2020,Banerjee2016}.

In contrast with conventional magnets, QSL appear in magnetic systems featuring
strong degrees of frustration, and are characterized by a quantum disordered
ground state\cite{Balents2010,Lee2008,Broholm2020,RevModPhys.89.025003,Savary_2016}. 
Interest in quantum-spin liquids has been fueled by their potential
emergent Majorana physics\cite{Kitaev2006} and their potential relation with
high-temperature superconductivity\cite{ANDERSON1987,PhysRevX.6.041007}. A variety of 
materials have been proposed as QSL candidates\cite{Han2012,Fu2015,Powell_2011, PhysRevX.9.031047,RevModPhys.88.041002, Takagi2019,
PhysRevLett.91.107001, Yamashita2008,
PhysRevB.77.104413,PhysRevLett.112.177201, PhysRevLett.98.107204,PhysRevB.100.144432,Bordelon2019}, yet it remains a remarkable challenge 
to unveil the nature of QSL and to experimentally identify them. 
Interestingly, magnetic van der Waals materials offer
new directions for the engineering and detection
of QSL, by exploiting the large flexibility of stacking and twisting
of moire systems.

\begin{figure}[t!]
\center
\includegraphics[width=\linewidth]{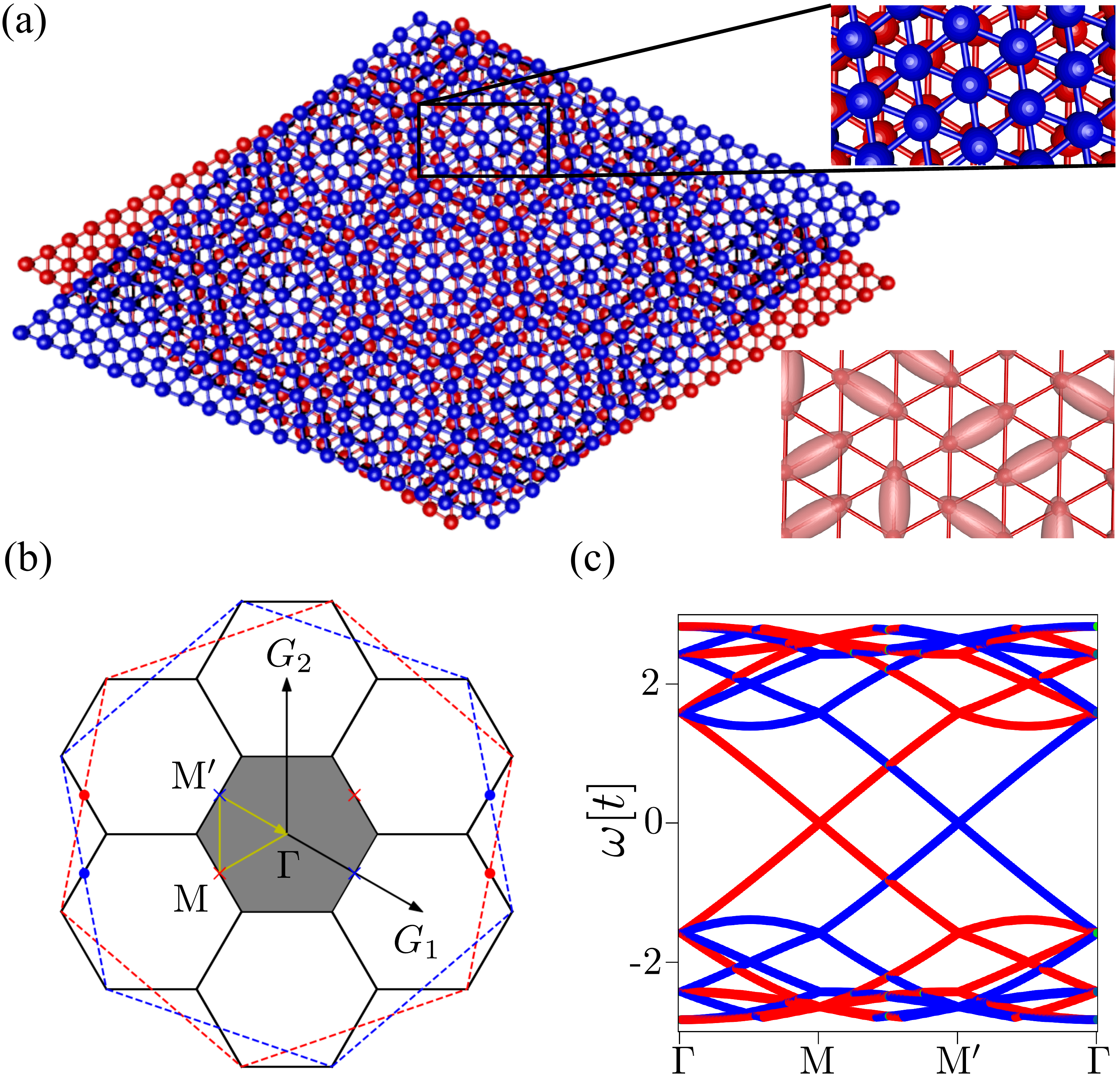}
\caption{(a) Sketch of a twisted bilayer Dirac QSL on a  triangular lattice.  (b)  The Moire Brillouin zone of the twisted QSL. The dashed red and blue hexagons represent the Brillouin zone of the bottom and top layers.
The red/blue circles (crosses) denote the location of the Dirac points in the original (folded) Brillouin zone (grey area). (c) Spinon bandstructure of the twisted Dirac QSL in
the decoupled limit for $\theta=22^\circ$, showing the two sets of decoupled bands in
the top (blue) and bottom (red) layers. }
\label{fig1}
\label{fig:fig1}
\end{figure}

Stacking van der Waals heterostructures yields electronic 
structures sensitive to the relative twisting between different layers.\cite{PhysRevB.82.121407,Bistritzer2011}. 
A paradigmatic example of these phenomena
is twisted bilayer graphene, where the emergence of flat bands
has lead to a variety of unconventional many-body states\cite{Cao2018,Cao2018super,Lu2019,Yankowitz2019,PhysRevLett.124.076801}.
Interestingly, twist engineering generically
provides a platform for
correlated phases with electrical
tunability\cite{Cao2020,wolf2020spontaneous,PhysRevLett.123.096802} and topologically nontrivial
electronic structures\cite{PhysRevB.99.235406,PhysRevLett.126.026801,PhysRevLett.124.106803,Serlin2019,PhysRevB.102.115127,Can2021}. 
The versatility offered by stacked van der Waals heterostructures motivates the search for 
analogous phenomena in the realm of van der Waals
magnets\cite{PhysRevB.101.245126,2020arXiv200811640Z} 
that can ultimately lead to novel spinon phenomena in
moire quantum spin-liquids.

In this Letter, we put forward twist engineering
in QSL van der Waals heterostructures
as a powerful knob to control spinon physics. We show that
twist engineering creates spinon flat bands at a specific twisting angle,
with a topological gap opening leading to in-gap spinon edge modes.
We show that the spinon spectra can be tuned by means of encapsulation
between van der Waals magnets, leading to dramatic changes in their
low energy spectra. Finally, we discuss how this exchange bias tuning
provides a spectroscopic electrical method
to characterize QSL states. Our results
put forward magnetic van der Waals heterostructures
formed by ferromagnets and QSLs as a tunable
platform to explore and probe spinon phenomena in moire systems.

For the sake of concreteness, in the following, we focus on a specific
gapless quantum spin liquid state on the triangular lattice,
structurally analogous to the one proposed for TaS$_2$.
For this sake, 
let us first briefly review
the physics of a single-layer QSL.
We start with a Heisenberg model
on a triangular lattice of the form
$
\mathcal{H}_0=\sum_{i,j}J_{ij}^{\mu\nu}S_i^{\mu}S_j^{\nu}
$
where $J_{ij}^{\mu\nu}$ are exchange coupling between spin components $\mu,\nu$ on sites $i,j$. 
The previous model is known to have a rich phase diagram and in particular
it supports QSL states such as those realized in TaS$_2$\cite{Law2017,PhysRevResearch.2.013099,manasvalero2020multiple}, TaSe$_2$\cite{Chen2020} and 
NaYbO$_2$\cite{PhysRevB.100.144432,Bordelon2019}.
We now focus on the regime
of the model yielding a QSL state
with a linear density of states (DOS), and in particular
the $U(1)$ Dirac spin 
liquid $\pi$-flux model\cite{PhysRevB.93.144411,PhysRevLett.123.207203}.
This state can be captured by performing a parton transformation
of the form
$
\mathbf{S}=\frac{1}{2}f^{\dag}_\alpha\mathbf{\sigma}_{\alpha\beta}f_\beta,
$
with $f^{\dag}_\alpha$ and $f_\alpha$ fermionic spinon operators and
$\sigma_{\alpha\beta}$ the spin Pauli matrices.
With the previous replacement, the Heisenberg model 
can be solved at the spinon mean-field level,
yielding a single-particle spinon Hamiltonian of the form
$
H_0 = t\sum_{\langle i,j\rangle}\chi_{ij}f^{\dag}_{i}f_j,
$
where $\chi_{ij}$ and $t$ are mean-field parameters. The $\pi$-flux model 
is defined by taking the mean-field solution
$\chi_{ij}$ hosting an associated staggered 0 and  $\pi$ fluxes in neighboring triangles.
We apply the gauge with real hoppings $\chi_{ij}=\pm1$ such that the system has time-reversal
symmetry. Under this gauge, the model has two Dirac cones
in the first Brillouin zone located at time-reversal invariant momenta\footnote{Here we choose a unit-cell containing $2\times2$ sites.}.

We move on to consider a twisted bilayer QSL as sketched in Fig.\ref{fig1}(a). 
We start from the parent Heisenberg Hamiltonian
for the twisted bilayer, that takes the form

\begin{eqnarray} 
\label{eq4}
\label{eq:heibi}
\mathcal{H}=\sum_{l,i,j}J_{\parallel,ij}^{\mu\nu}S_{i,l}^{\mu}S_{j,l}^{\nu}+\sum_{i,j}J_{\perp,ij}^{\mu\nu}S_{i,1}^{\mu}S_{j,2}^{\nu},
\end{eqnarray}
where $l$ labels the two layers, and $J_{\parallel,ij}^{\mu\nu}$ and $J_{\perp,ij}^{\mu\nu}$ denote intra- and inter-layer spin exchange, respectively. In the regime $J_{\parallel,ij}^{\mu\nu}\gg J_{\perp,ij}^{\mu\nu}$, the ground state
of the system will consist on two coupled $U(1)$ QSL states. 
Therefore, we take as the mean-field solution for each layer 
the spinon $\pi$-flux model, with an effective inter-layer spinon coupling from inter-layer spin exchange:

\begin{figure}[t!]
\center
\includegraphics[width=\linewidth]{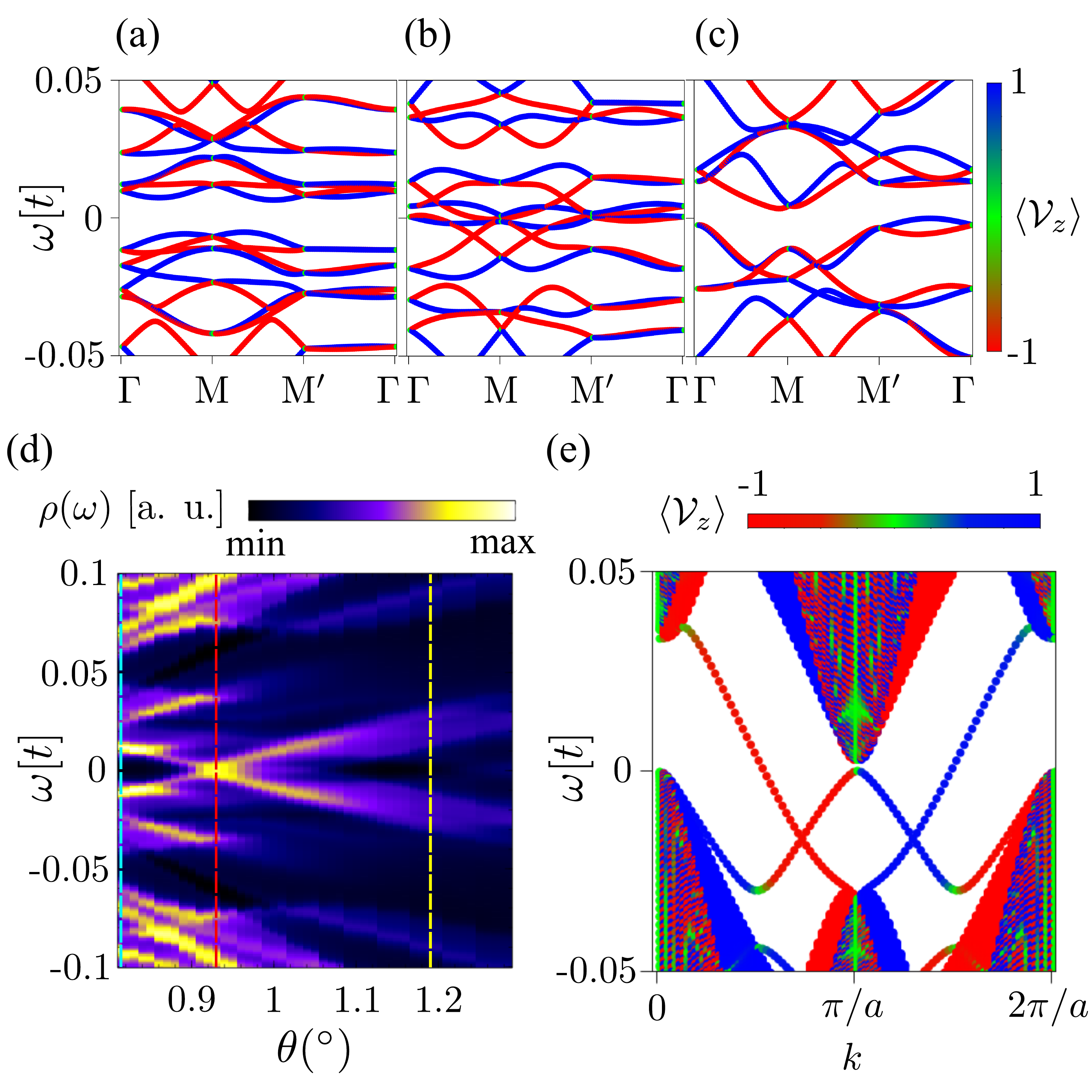}
\caption{(a-c) Spinon bandstructure of the twisted Dirac QSL at different twisting angles $\theta=0.81^\circ,0.93^\circ,1.20^\circ$ (dashed lines in panel (d)), respectively. (d) Spinon DOS $\rho(\omega)$ of the twisted Dirac QSL near the flat band twisting angle $\theta=0.93^\circ$. (e) Spinon bandstructure of a twisted QSL nanoribbon at $\theta=3.5^\circ$.}
\label{fig2}
\end{figure}

\begin{eqnarray} \label{eq5}
H=t\sum_{l,\langle i,j\rangle}\chi_{l,ij}f_{i,l}^{\dag}f_{j,l}+\sum_{i,j}t_{\perp,ij}(f_{i,1}^{\dag}f_{j,2}+h.c.),
\end{eqnarray}
where $t, \chi_{l,ij}$ and $t_{\perp,ij}$ are mean-field parameters
that can be derived analogously from a mean-field replacement in Eq. \eqref{eq:heibi}.
The mean-field parameter $t_{\perp,ij}$ will depend
on the relative distance between sites $i$ and $j$, inherited from the
parent Heisenberg coupling
$J_{\perp,ij}^{\mu\nu}$ in Eq. \eqref{eq4}. 
We take a functional form for the
interlayer coupling as\cite{PhysRevB.92.075402} $t_{\perp,ij}=t_{\perp,0}\frac{d^2}{r^2_{ij}}e^{-\lambda(r_{ij}-d)}$, where $d$
is the inter-layer distance, $r_{ij}$ is the distance between sites $i$ and $j$, $\lambda$ is
the parameter that controls the decay of the inter-layer coupling, and $t_{\perp,0}$ is
the largest possible inter-layer coupling realized at $r_{ij}=d$.
In the following we take $t_{\perp,0}=0.36t$, $\lambda=10/a$, and $d=a$, where $a$ is the lattice constant of the triangular lattice.
From the computational point of view, we will use the twist scaling relation
for computational convenience\cite{PhysRevLett.114.036601, PhysRevLett.119.107201},
and we compute the valley expectation
$\langle \mathcal{V}_z\rangle=\pm 1$
by means of the valley operator\cite{PhysRevLett.120.086603,wolf2020spontaneous,Soriano2020,PhysRevB.99.245118,PhysRevResearch.2.033357}.

The $\pi$-flux hoppings $\chi_{l,ij}$ are subject to a $U(1)$ degree of freedom for 
each layer, respectively. However, the gauge difference between the two layers determines 
the relative position of Dirac cones of the two layers in reciprocal space\cite{PhysRevLett.98.117205}. 
As a result, the momentum difference between Dirac cones of the two layers, $\Delta \mathbf{k}$, 
can be either large or small. When $|\Delta \mathbf{k}|\gg |1/\mathbf{R}|$, where $\mathbf{R}$ 
is the periodicity of $t_{\perp,ij}$ in real space, the Dirac cones are almost decoupled. 
In such case, the impact of $t_{\perp,ij}$ is small on low energy physics,
keeping the two layers effectively decoupled.
In contrast, when $|\Delta \mathbf{k}|\ll |1/\mathbf{R}|$,  $t_{\perp,ij}$ leads
to significant coupling between the Dirac cones of different layers.
From the energetic point of view, this gauge configuration
couples the two layers and therefore will
lower the many-body energy
through spinon hybridization.
With this gauge choice, the Moire Brillouin zone is shown in Fig.\ref{fig1}(b), 
with two Dirac cones at time-reversal invariant momenta M and M$'$.
The spinon dispersion in the decoupled
limit for twisting angle $\theta \approx 22^\circ$ is shown in Fig.\ref{fig1}(c),
where the two spinon Dirac cones of the two decoupled layers are observed.

Let us now move on to the case in which the
two QSL are coupled through the interlayer
exchange coupling.
In this situation,
we find that the
interlayer coupling
leads to a gap opening  in the Dirac cones
of the bilayer QSL upon twisting.
This gap opening stemming from the twist is similar to 
the case of twisted double bilayer 
graphene\cite{PhysRevB.99.235406} (Fig.\ref{fig2}(a-c)),
and stems from the broken $C_{2z}T$
symmetry\cite{PhysRevX.9.021013,PhysRevX.8.031089} of the 
effective model.
Besides the gap opening, we
also observe
the emergence
of spinon flat bands at a specific fine tuned
twisting angle
$\theta/(t_{\perp}/t) \approx 2.6^\circ$,
which for $t_{\perp}=0.36t$ appears at
$\theta=0.93^\circ$ (Fig.\ref{fig2}(d)), 
similar to other twisted Dirac materials\cite{PhysRevB.82.121407,Bistritzer2011}.

Interestingly, the emergence of a gap opening driven by the twist has been
shown to give rise to topological states in van der Waals heterostructures based on graphene\cite{PhysRevLett.126.056401,PhysRevX.9.031021,PhysRevB.99.235417}.
In particular, we find that the gap opening in the bilayer QSL has an associated
valley Chern number of $2$, giving rise to two counterpropagating
channels at each edge with opposite valley polarization\cite{Prada2011,Zhang2013,Wright2011,PhysRevB.88.121408,Rickhaus2018}.
The previous phenomenology can be explicitly demonstrated by computing
the spinon band structure of a
twisted
QSL nanoribbon at twisting angle $\theta=3.5^\circ$ (Fig.\ref{fig2}(e)). 
In particular, it is clearly seen the emergence
of in-gap
valley-polarized topological edge modes, 
associated with the topological valley Hall quantum spin-liquid state. 
We note that the topological edge modes are protected by the 
approximate valley charge conservation, and therefore perturbations
giving rise to strong intervalley scattering can
lead to intervalley mixing between topological spinon edge states\cite{PhysRevLett.97.196804,PhysRevLett.102.236805,PhysRevB.99.245118,PhysRevResearch.2.033357,PhysRevB.88.121408}.

\begin{figure}[t!]
\center
\includegraphics[width=\linewidth]{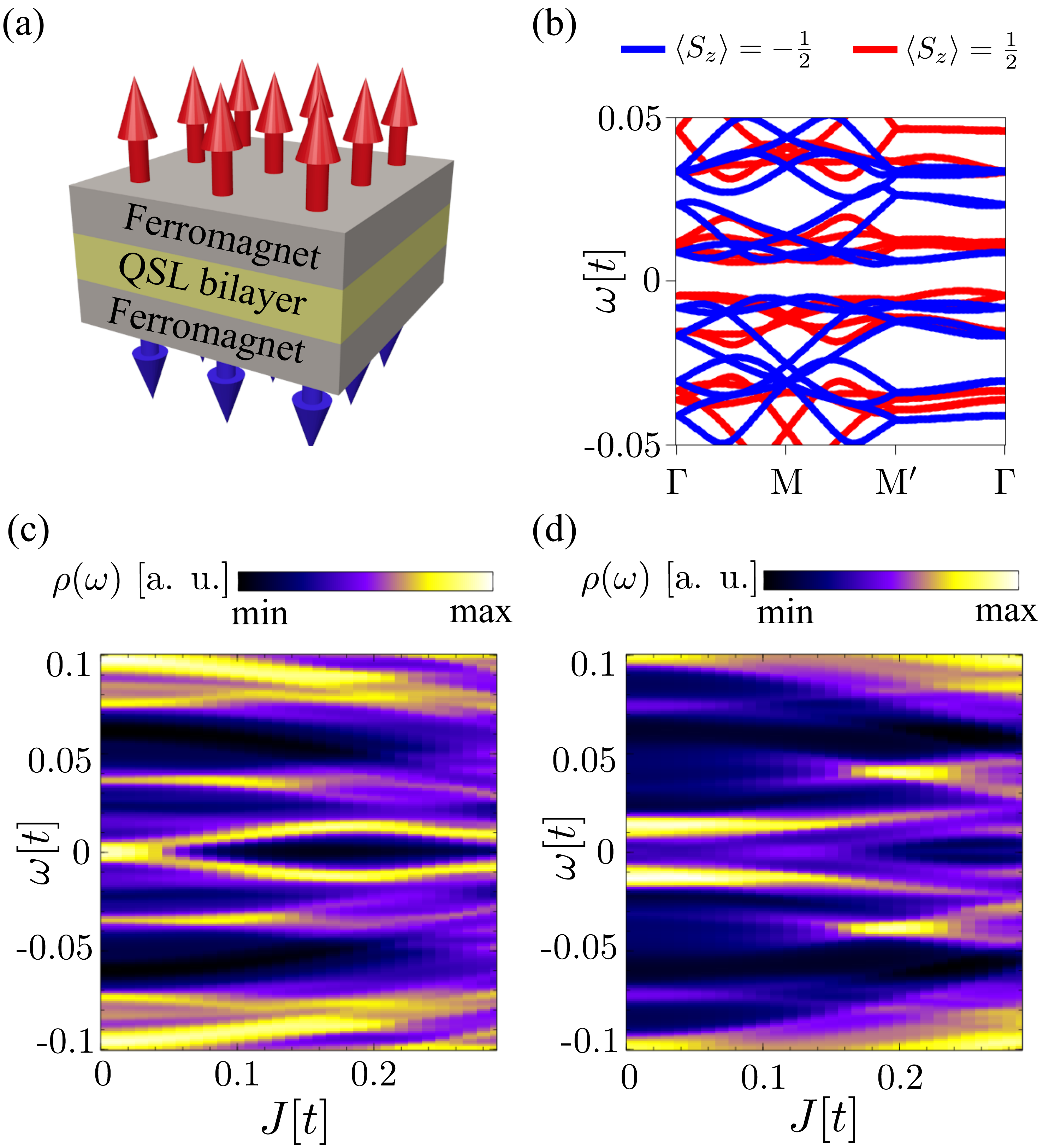}
\caption{(a) Sketch of the twisted bilayer Dirac QSL encapsulated by ferromagnets. (b) Spinon bandstructure of the twisted Dirac QSL at $\theta=0.93^\circ$ and inter-layer spinon bias $J=0.1t$. (c,d) Spinon DOS of the twisted bilayer Dirac QSL at (c) $\theta=0.93^\circ$ and (d) $\theta=1.05^\circ$ with different inter-layer spinon bias $J$.}
\label{fig3}
\end{figure}

After considering the emergent spinon spectra driven by the twist between the two QSL layers, we now move
on to show how such emergent moire bands can be controlled by a magnetic encapsulation.
From the material science point of view, in the following
we will consider that the QSL bilayer is sandwiched between two ferromagnetic
insulators, for which both CrBr$_3$\cite{ghazaryan2018magnon,kezilebieke2020electronic} and CrCl$_3$ would be suitable candidates. 
The top and bottom ferromagnets
are expected to be antiferromagnetically aligned
through a superexchange mechanism\cite{PhysRev.79.350},
as shown in Fig. \ref{fig3}(a). 
To study the impact on the QSL state, we now integrate out the degrees of freedom
of the ferromagnet, and consider their impact on the QSL Hamiltonian.
The magnetic encapsulation
yields an exchange proximity term in the Hamiltonian of the QSL bilayer, analogous
to the exchange terms proposed
for other van der Waals materials proximized to ferromagnets\cite{PhysRevLett.121.067701,PhysRevLett.118.187201,Wei2016,PhysRevLett.114.016603,wolf2020spontaneous}

\begin{eqnarray} \label{eq6}
\mathcal{H}'=\mathcal{H}+\sum_{i,\mu}\mathcal{J}_{\perp}S_{i,1}^{\mu}\mathcal{M}_{1}^{\mu}
+\sum_{i,\mu}\mathcal{J}_{\perp}S_{i,2}^{\mu}\mathcal{M}_{2}^{\mu},
\end{eqnarray}
where $\mathcal{J}_{\perp}$ denotes the exchange interaction between spin in the QSL
$S_{i,l}^{\mu}$ and the magnetic moment of the ferromagnets
$\mathcal{M}_{l}^{\mu}$, with $l=1,2$ 
labelling the two different magnets and QSL layers.
We consider sufficiently small $\mathcal{J}_{\perp}$ 
that does not lead to many-body
reconstruction. 
In such case, the mean-field solution of Eq. (\ref{eq5}) remains, and the 
effect of the ferromagnets can be projected onto the 
spinon mean-field Hamiltonian as

\begin{eqnarray} \label{eq7}
H'=H+\frac{1}{2}\sum_{i,\mu,l,s,s'}{\mathcal{J}}_{\perp}\sigma^{\mu}_{s,s'}
\mathcal{M}_{l}^{\mu}f^{\dag}_{i,s,l}f_{i,s',l}.
\end{eqnarray}

In the case of the antiferromagnetic alignment as
depicted in Fig. \ref{fig3}(a), the magnetic encapsulation creates
an effective spin-dependent inter-layer spinon bias
$J=\mathcal{J}_\perp (\mathcal{M}^z_1 - \mathcal{M}^z_2)$ 
on the twisted Dirac QSL\cite{PhysRevLett.121.067701,wolf2020spontaneous,Soriano2020}. 
The effective inter-layer spinon bias $J$ induced by proximity exchange fields has a dramatic impact on
the low energy spinon bandstructure of the twisted Dirac QSL. Due to the broken mirror symmetry of the twisted QSL\footnote{The mirror symmetry $M_x$($M_y$) is present if the system is invariant when mirrored by $x$ ($y$) plane and then exchanging the top and bottom layers.}, the exchange bias causes
spin-splitting in the spinon bandstructure in reciprocal space (Fig.\ref{fig3}(b)). 
At the flat band twisting angle, the gap at Fermi
level drastically increases with $J$ (Fig.\ref{fig3}(c)). When the twisting angle is not
the flat one, the spinon DOS gets
substantially modified at larger exchange couplings
(Fig.\ref{fig3}(d)). 

\begin{figure}[t!]
\center
\includegraphics[width=\linewidth]{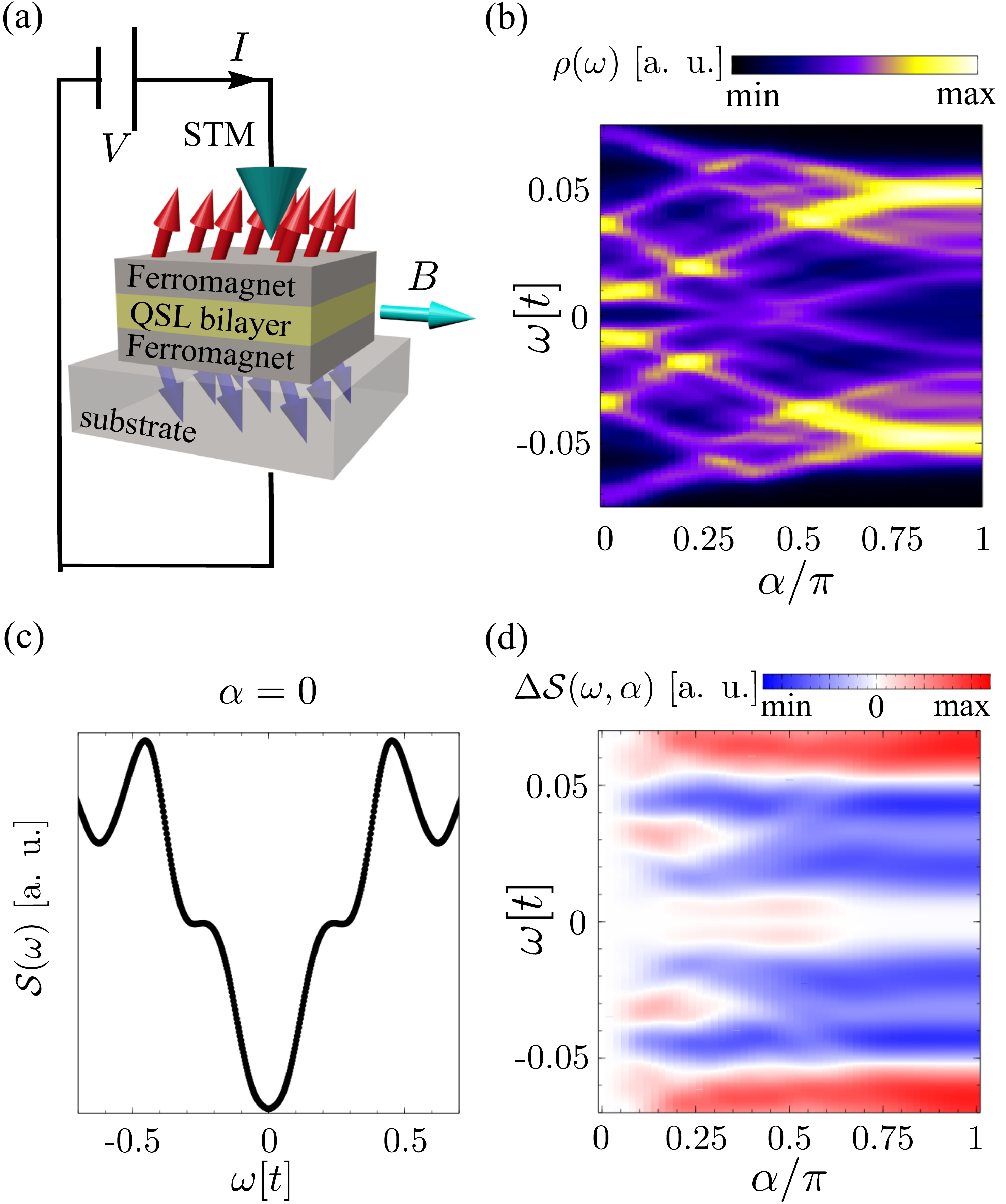}
\caption{(a) Sketch of the experimental setup for probing QSL via inelastic spectroscopy. (b) Spinon DOS of the twisted QSL at $\theta=0.93^\circ$ and $J=0.1t$ under different $\alpha(B)$. (c) Spin structure factor $\mathcal{S}(\omega)$ with $\alpha=0$. (d) Change in spin structure factor $\Delta \mathcal{S}(\omega,\alpha)$ for different $\alpha$.}
\label{fig4}
\end{figure}

We now move on to discuss how the impact of the exchange bias can be used
for the detection of the spinon spectra.
Scanning tunnel spectroscopy can be exploited to detect spin excitations
by means of inelastic transport\cite{PhysRevLett.125.267206,ghazaryan2018magnon,PhysRevResearch.2.033466,Klein1218}.
We now consider a similar setup, in which
the magnetically encapsulated QSL bilayer
is explored by means of vertical transport
with STM as shown in Fig. \ref{fig4}(a). 
From the experimental point of view, spinon states will emerge in the inelastic contribution to the
current, and will appear as peaks in
$\text{d}^2I/\text{d}V^2$ when phonon contributions are neglected\footnote{Spinon energy scale 
is smaller than the magnon gap in ferromagnets, allowing us to neglect the contribution from inelastic
scattering with magnons.}.
In such case, the measured $\text{d}^2I/\text{d}V^2$ is proportional to the 
spin structure factor $\mathcal{S}(\omega)$, the latter being a convolution of spinon
DOS\cite{spinelli2014imaging,PhysRevResearch.2.033466}: 
$\mathcal{S}(\omega)=\int \text{d}\omega_1\text{d}\omega_2\frac{\rho(\omega_1)\rho(\omega_2)}{\omega+\omega_1-\omega_2+i0^{+}}(f(\omega_1)-f(\omega_2))$ where $f(\omega)$ is the Fermi-Dirac distribution.

To reveal the impact of the magnetic encapsulation,
we now consider the change of the signal with respect to an in-plane magnetic field $B$,
that is used to control
the direction
of magnetism in the magnets. The magnetic field will tune the angle between magnetization of the two magnets
from $\pi$ to $\pi-\alpha(B)$, and modifies spinon DOS in the QSL bilayer due to proximity effect. 
For
twisting angle $\theta=0.93^\circ$, and effective exchange bias $J=0.1t$, the spinon DOS 
under different $\alpha(B)$ is shown in Fig. \ref{fig4}(b). The modified spinon DOS exhibits peaks at
different frequencies than the original one. 
Taking the specific case of antiferromagnetic alignment $\alpha=0$,
it is seen that the spin structure factor has peak structures, inherited
from the peaks in the spinon DOS (Fig.\ref{fig4}(c)). 
Due to the modification of the spinon DOS with the field, an analogous
effect is expected in the d$^2$I/dV$^2$.
For this sake, we now compute
the change in the d$^2$I/dV$^2$ as a function of the magnetic field,
defined as
$\Delta \mathcal{S}(\omega,\alpha)=\mathcal{S}(\omega,\alpha)-\mathcal{S}(\omega,0)$,
and shown in Fig. \ref{fig4}(d).
In particular, we see in Fig. \ref{fig4}(d) 
the existence of deeps and peaks in the differential
d$^2$I/dV$^2$, manifesting from the dramatic change of
spinon DOS with the magnetic field of Fig. \ref{fig4}(b).

Finally, we comment on specific quantitative aspects of our proposal relevant for experiments.
First, in our manuscript, we have considered exchange couplings between the two monolayer QSL
on the order of $J_\perp \approx 0.3 J_{||}$. The specific prediction of such
exchange coupling should be performing via first principle methods for the specific materials
considered\cite{Sivadas2018,Soriano2019},
and could ultimately be controlled with pressure\cite{Song2019,Li2019}.
For our phenomenology, a change in the exchange coupling just drifts the physics towards
bigger or smaller angles, yet without qualitatively
changing the overall behavior\cite{PhysRevLett.99.256802}.
Second, when exchange proximity is considered, the exchange proximity must be smaller
than the intralayer exchange to not perturb the QSL ground state.
Third, in order to tilt the direction of the magnets by an external magnetic field, yet without
breaking the QSL ground state, a soft magnetic axis is
preferred. As a reference,
taking CrBr$_3$ as the ferromagnet, 
the anisotropy energy can be overcome
with
a magnetic field of around $1$ T\cite{PhysRevMaterials.2.024004}, 
whose Zeeman energy scale of 0.04 meV is not expected
to perturb QSL with exchange constants on the order of 5 meV.
The application of a magnetic field is also expected to affect the magnons of the ferromagnetic encapsulation,
yet those states will contribute with a uniform background to the
d$^2$I/dV$^2$ signal that can be substracted\cite{Klein1218,ghazaryan2018magnon}.
Finally, although our analysis has focused on a Dirac QSL, analogous calculations
can be performed with other QSL ground states, such as the gaped QSL of RuCl$_3$\cite{Kitaev2006}.  

To summarize, we have shown that a van der Waals heterostructure based on a magnetically
encapsulated bilayer QSL allows designing controllable
spinon physics, and ultimately detecting moire spinons. 
We showed that a twisted bilayer QSL gives
rise to a topological gap opening
purely driven by the twist, and in fine-tuned regimes
spinon flat bands. 
Furthermore, we showed that encapsulating the twisted QSL bilayer with ferromagnets produces 
exchange bias via proximity effect, which allows to magnetically control the spinon bandstructure. 
Based on such magnetic tunability, we proposed an experimental identification of QSL phases, 
utilizing field-controlled magnetism in magnetic van der Waals materials as well as inelastic spectroscopy.
Our results put forward twist engineering as a means of exploiting exotic spinon phenomena 
in quantum spin-liquids, highlighting the versatility of magnetic van der Waals heterostructures
to explore emergent spinon phenomena.

\begin{acknowledgments}
We acknowledge
the computational resources provided by
the Aalto Science-IT project,
and the
financial support from the
Academy of Finland Projects No.
331342 and No. 336243.
We thank P. Liljeroth, 
S. Kezilebieke, V. Vaňo,
and S. Ganguli for fruitful discussions.
\end{acknowledgments}

\begin{appendix}

\end{appendix}

\bibliographystyle{apsrev4-1-etal-title}
\bibliography{QSL}
\bibliographystyle{apsrev4-1}

\end{document}